\documentclass[%
aps,
superscriptaddress,nofootinbib,
reprint,
]{revtex4-2}

\usepackage{xcolor}
\usepackage{graphicx}
\graphicspath{{./fig/}}
\usepackage{amsmath,amssymb,amsfonts,mathrsfs,bm}
\usepackage[%
colorlinks=true,
linkcolor=blue,
citecolor=blue,
]{hyperref}

\begin{document}
	
\title{Explaining Pulsar Timing Array Observations with Primordial Gravitational Waves in Parity-Violating Gravity}

\author{Chengjie Fu}
\email[]{fucj@ahnu.edu.cn}
\affiliation{Department of Physics, Anhui Normal University, Wuhu, Anhui 241002, China}

\author{Jing Liu}
\email[]{liujing@ucas.ac.cn}
\affiliation{International Centre for Theoretical Physics Asia-Pacific, University of Chinese Academy of Sciences, Beijing 100190, China}
\affiliation{Taiji Laboratory for Gravitational Wave Universe (Beijing/Hangzhou), University of Chinese Academy of Sciences, Beijing 100049, China} 

\author{Xing-Yu Yang}
\email[]{xingyuyang@kias.re.kr}
\affiliation{Quantum Universe Center (QUC), Korea Institute for Advanced Study, Seoul 02455, Republic of Korea}

\author{Wang-Wei Yu}
\email[]{yuwangwei@mail.itp.ac.cn}
\affiliation{CAS Key Laboratory of Theoretical Physics, Institute of Theoretical Physics, Chinese Academy of Sciences, Beijing 100190, China}
\affiliation{School of Physical Sciences, University of Chinese Academy of Sciences, Beijing 100049, China} 

\author{Yawen Zhang}
\affiliation{Department of Physics, Anhui Normal University, Wuhu, Anhui 241002, China}

\begin{abstract}
The pulsar timing array (PTA) collaborations have recently suggested the presence of a gravitational wave background at nano-Hertz frequencies. In this paper, we explore potential inflationary interpretation of this signal within the context of a simple and health parity-violating gravity model termed the Nieh-Yan modified Teleparallel Gravity. Through this model, two inflationary scenarios are evaluated, both yielding significant polarized primordial gravitational waves (PGWs) that align well with the results from PTA observations. Furthermore, the resulting PGWs can display strong circular polarization and significant anisotropies in the PTA frequency band, which are distinct features to be verified by observations of both PTA and the cosmic microwave background. 
The detection of such a distinctive background of PGWs is expected to provide strong evidence supporting our scenarios and insights into inflationary dynamics and gravity theory.

\end{abstract}

\maketitle

\section{Introduction}
Inflation is a successful model of the very early Universe that solves the flatness and horizon problems and sets the initial conditions of big bang cosmology. During inflation, the Universe undergoes a period of exponential expansion, during which curvature perturbations from quantum fluctuations of the scalar modes successfully explain the temperature anisotropies in the cosmic microwave background (CMB) and seed the large-scale structure. Primordial gravitational waves (PGWs) are unique prediction of inflation, and detecting them has been considered as a smoking gun for inflation. However, PGWs have yet to be discovered, despite decades of searches in B-mode polarizations in the CMB. The latest constraint on the tensor-to-scalar ratio is obtained by the combination of Planck and BICEP/Keck, with $r < 0.036$ at the 95\% confidence level~\cite{BICEP:2021xfz}. Apart from observations at the CMB scales, other gravitational wave observation programs, such as the LIGO-Virgo-KAGRA collaboration \cite{KAGRA:2021kbb}, LISA/Taiji/Tianqin \cite{Barausse:2020rsu,Ruan:2018tsw,TianQin:2015yph}, and pulsar timing array (PTA) experiments~\cite{NANOGrav:2023gor,Reardon:2023gzh,EPTA:2023fyk,Xu:2023wog}, offer opportunities to observe PGWs at small scales with modified gravity models or non-standard thermal histories of the Universe.

Recently, the PTA experiments, including NANOGrav \cite{NANOGrav:2023gor}, PPTA \cite{Reardon:2023gzh}, EPTA \cite{EPTA:2023fyk}, and CPTA \cite{Xu:2023wog}, collectively reported an exciting discovery of nano-Hertz stochastic gravitational wave background (SGWB). Beyond the standard astrophysical interpretation in terms of inspiraling supermassive black hole binaries \cite{NANOGrav:2023hfp,Ellis:2023dgf,Bi:2023tib,Cannizzaro:2023mgc}, an SGWB stemming from cosmological sources may serve as a potential explanation of the observed signal. This breakthrough promises transformative insights into early universe cosmology. Cosmological SGWBs are expected to originate from vacuum fluctuations or particle production during inflation \cite{Bartolo:2016ami,Mylova:2018yap,Odintsov:2021kup,Oikonomou:2022ijs,Zhu:2022dfq,Sorbo:2011rz,Cook:2011hg,Barnaby:2011qe,Adshead:2013qp,Namba:2015gja,Domcke:2016bkh,Peloso:2016gqs,Ozsoy:2020kat,DAmico:2021vka,Cai:2021wzd}, as well as from large primordial scalar perturbations produced during inflation~\cite{Di:2017ndc,Cai:2018dig,Bartolo:2018evs,Inomata:2018epa,Cai:2019amo,Cai:2019jah,Yuan:2019udt,Fu:2019ttf,Fu:2019vqc,Cai:2019bmk,Lin:2020goi,Pi:2020otn,Vaskonen:2020lbd,DeLuca:2020agl,Inomata:2020xad,Yi:2020cut,Gao:2020tsa,Domenech:2021ztg}, cosmological first-order phase transitions ~\cite{Kamionkowski:1993fg,Caprini:2007xq,Cutting:2018tjt,Hindmarsh:2015qta,Guo:2020grp}, and topological defects such as cosmic strings ~\cite{Vilenkin:1984ib,Damour:2000wa,Auclair:2019wcv} and domain walls ~\cite{Vilenkin:1984ib,Hiramatsu:2013qaa,Saikawa:2017hiv}. A comprehensive review can be found in \cite{Caprini:2018mtu}. 
The SGWBs originating from these assorted sources have been extensively examined as plausible interpretations of the PTA signal  \cite{NANOGrav:2023hvm,Bian:2023dnv,Vagnozzi:2023lwo,Franciolini:2023pbf,Oikonomou:2023qfz,Wang:2023ost,Liu:2023ymk,Liu:2023pau,Unal:2023srk,Gouttenoire:2023bqy,HosseiniMansoori:2023mqh,He:2023ado,Ellis:2023oxs,Yi:2023npi,Choudhury:2023kam,Choudhury:2023wrm,Bhattacharya:2023ysp}.
In this study, we concentrate on the vacuum fluctuations during inflation, namely PGWs, exploring the inflationary explanation of the PTA signal within the context of the parity-violating gravity theory.

Chern-Simons theory~\cite{Jackiw:2003pm,Alexander:2009tp} is the most widely discussed parity-violating theory of modified gravity. It has been used to generate polarized PGWs~\cite{Lue:1998mq,Satoh:2007gn,Alexander:2004us,Alexander:2016hxk,Qiao:2019hkz,Odintsov:2022hxu}. However, Chern-Simons theory suffers from ghost instabilities induced by higher-order derivative terms, which have been addressed in more complex models~\cite{Crisostomi:2017ugk,Gao:2019liu,Zhu:2022uoq}.

In this work, we investigate a novel modified gravity model that allows for the amplification and observation of PGWs with strong circular polarization, and even large anisotropies. We consider a simple and healthy gravity theory with parity violation~\cite{Li:2020xjt,Li:2021wij}, where the parity symmetry is violated by introducing the scalar field coupled Nieh-Yan term into the teleparallel equivalent of general relativity.
Given teleparallel equivalent of general relativity is dynamically equivalent to general relativity (GR), this is equivalently making a parity-violating extension to GR. The first constraint on this model was conducted in \cite{Wu:2021ndf} through utilizing observational data from LIGO-Virgo GW events. Ref. \cite{Bombacigno:2022naf} studied the behavior of GWs from this parity-violating gravity model propagating in a medium of collisionless particles.
The implications that the Nieh-Yan modified Teleparallel Gravity (NYmTG) model holds in the context of inflation have been scrutinized in prior work \cite{Cai:2021uup}, which suggests that the anticipated energy spectrum of PGW signal exhibits a significant bump at frequencies nestled within the range perceptible by prospective GW experiments, surpassing their sensitivity curves. Consequently, it is plausible to infer that the emergent PGWs within the NYmTG model could provide a satisfactory explanation for the PTA signal.

In our scenario, the coupled scalar field acts as either an inflaton or a spectator field with its dynamics during inflation triggering the amplification of PGWs. Such amplified GWs can adequately explain the PTA data.
Since the gravitational sources after inflation are constrained by other observables, such as upper bounds on small-scale curvature perturbations~\cite{Liu:2022lvz,Ramberg:2022irf,Liu:2023hte} and observations of GWs at other frequency bands~\cite{LIGOScientific:2021nrg,Romero:2021kby}, PGWs as a product of inflation can safely avoid these constraints and become a promising candidate for cosmological SGWBs to explain the PTA signal.
Moreover, this model predicts a characteristic SGWB with simultaneously large anisotropies and large polarization. 
Since the polarization of an isotropic SGWB cannot be detected by PTA experiments \cite{Kato:2015bye,Belgacem:2020nda}, this work provides a good opportunity to test both inflation and parity-violating modified gravity model with PTA experiments in the near future.

Throughout the paper, we will use the units $c=\hbar=1$, the reduced Planck mass $M_{\rm P}=1/\sqrt{8\pi G}=1$, and the convention for the metric signature $(+,-,-,-)$.

\section{Model}
We briefly revisit the foundational formulas intrinsic to the NYmTG model, which is characterized by the following action \cite{Li:2020xjt,Li:2021wij},
\begin{align}\label{action}
\notag
S  &= \int d^4 x \sqrt{-g}\left[-\frac{R}{2} + \frac{\alpha\phi}{4} \mathcal{T}_{A\mu\nu} \tilde{\mathcal{T}}^{A \mu \nu} + \frac{1}{2}\nabla_\mu \phi\nabla^\mu \phi -V(\phi) \right]\\
&~~~+ S_{\rm {other}},
\end{align}
where $R$ is the Ricci scalar, $\tilde{\mathcal{T}}^{A \mu \nu}=(1/2)\varepsilon^{\mu\nu\rho\sigma}\mathcal{T}^A_{~~\rho\sigma}$ represents the dual of the torsion two form $\mathcal{T}^A_{~~\mu\nu}$ with $\varepsilon^{\mu\nu\rho\sigma}$ being the Levi-Civita tensor, and $\alpha$ is the coupling constant. Here, the $\phi$ field emerges as a dynamical scalar field, and the action $S_{\rm {other}}$ describes an additional canonical field that is minimally coupled to gravity. With the spatially flat FLRW metric, the background evolution is identical to that in GR, governed by the following equations:
\begin{subequations}\label{eq:BGeq}
\begin{align}
3H^2 = \frac{1}{2}\dot\phi^2 + V(\phi) + \rho,\\
\ddot\phi+3H\dot\phi +\frac{dV}{d\phi}=0,\\
\dot\rho + 3H(\rho + p)=0,
\end{align}
\end{subequations}
where $H$ is the Hubble parameter and a dot denotes the derivative with respect to the cosmic time. Here, $\rho$ and $p$ denote the energy density and pressure, respectively, associated with the additional field. It is worth noting that in this model, gauge invariant scalar perturbations corresponding to $\phi$ vanishes completely due to the coupling of the $\phi$ field with the Nieh-Yan term. Consequently, if the $\phi$ field serves as the inflaton, it cannot generate the primordial scalar perturbations. The curvaton scenario, which explains the origin of primordial scalar perturbations in the context of the $\phi$ field acting as an inflaton, has been thoroughly analyzed in~\cite{Cai:2021uup}.

In this model, the tensor perturbations exhibit the phenomenon of velocity birefringence, obeying the following equation of motion in the momentum space,
\begin{align}\label{eq:EoM_h_k}
\ddot h^A_k + 3H \dot h^A_k + \frac{k}{a} \left(\frac{k}{a} + \lambda_A \alpha \dot \phi  \right)h^A_k = 0,
\end{align}
where $A=R(L)$ denotes the right(left)-handed polarization with $\lambda_R=1$ and $\lambda_L=-1$. A salient characteristic of the tensor perturbation within this context is the occurrence of tachyonic instability in one of its two polarization states under particular circumstances. For instance, during inflation, there invariably exists a polarization state where the frequency squared of the corresponding modes, denoted by $(k/a)(k/a+\lambda_A \alpha \dot{\phi})$, becomes negative as the physical wave number $k/a$ evolves to be less than $|\alpha\dot{\phi}|$. If the condition $|\alpha\dot{\phi}|>\mathcal{O}(H)$ is met, certain mode functions $h_k^A$ will experience a tachyonic instability within the horizon, instigating an exponential amplification of their amplitude~\cite{Cai:2021uup}.

By incorporating the above distinctive characteristic into specific inflationary scenarios, we can expect a PGW interpretation of the PTA signal. In this paper, we consider two distinct scenarios: 1) the field $\phi$ acts as the inflaton while $S_{\text{other}}$ represents a curvaton field, as illustrated in~\cite{Cai:2021uup}; 2) the field $\phi$ serves as the spectator field while $S_{\text{other}}$ describes an inflaton. 

\begin{figure}[htbp]
	\centering
	\includegraphics[width=0.9\columnwidth]{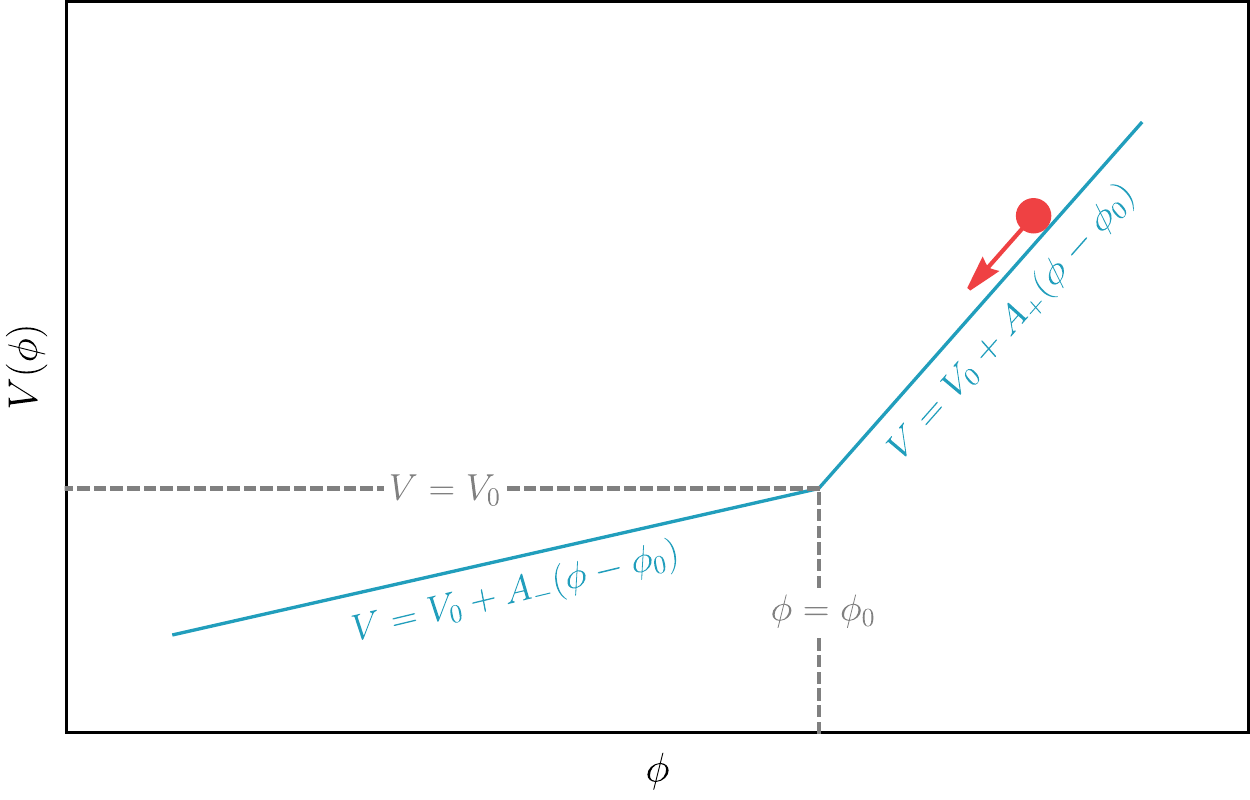}\\
    \includegraphics[width=0.9\columnwidth]{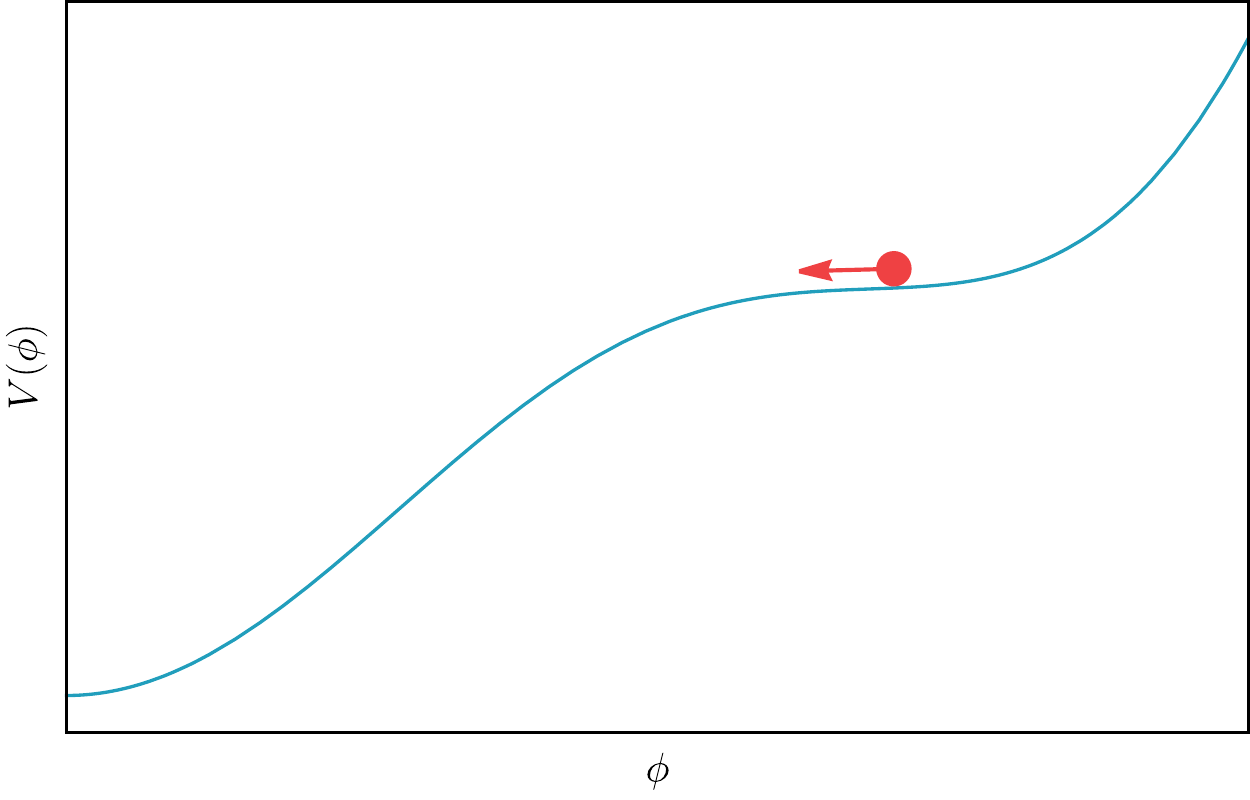}
	\caption{Schematic diagrams of the Starobinsky’s linear potential (top) and the axion potential (bottom).}
    \label{fig:schematic} 
\end{figure}

For the first scenario, we postulate that the inflaton $\phi$ is endowed with Starobinsky’s linear potential, expressed as~\cite{Starobinsky:1992ts},
\begin{align}
V(\phi)=\left\{
\begin{array}{cl}
V_0 + A_+(\phi-\phi_0), \qquad \text{for}~\phi>\phi_0, \\
V_0 + A_-(\phi-\phi_0), \qquad \text{for}~\phi\leq \phi_0, \\
\end{array} \right.
\end{align}
depicted in the top panel of Fig.~\ref{fig:schematic}.
Here, $V_0$ determines the inflationary energy scale, while the parameters $A_+$ and $A_-$ control the slopes of two linear potentials. The amplification of the small-scale primordial scalar spectrum, along with the formation of primordial black holes in the canonical single-field inflation with this potential, has been comprehensively investigated in~\cite{Pi:2022zxs}.
However, in this paper, we can avoid discussing these phenomena due to the vanishing perturbations in the $\phi$ field.

For the second scenario, we choose the axion potential~\cite{Kobayashi:2015aaa} to describe the field $\phi$, which is given by
\begin{align}\label{axion_pot}
    V(\phi) = \frac{1}{2}m^2\phi^2 + \Lambda^4 \frac{\phi}{\sigma}\sin\left( \frac{\phi}{\sigma} \right),
\end{align}
as shown in the bottom panel of Fig.~\ref{fig:schematic}. This choice is the same as in~\cite{Cai:2021uup}, but the difference lies in our treatment of the field $\phi$ as a spectator field rather than as the inflaton. In the following section, we will explore the phenomenological implications of combining the NYmTG model with these two inflationary paradigms.

\section{Results}

We perform numerical analyses to extract the PGW predictions arising from the NYmTG model within the previously mentioned inflationary contexts. By numerically solving the coupled set of background equations given in Eqs.~\eqref{eq:BGeq} and the tensor perturbation equation presented in Eq.~\eqref{eq:EoM_h_k}, we can obtain the power spectrum of the tensor perturbations, $\mathcal{P}_h(k) =\sum_{A=R,L}k^3|h_k^A|/(2\pi^2)$. Subsequently, we deduce the current energy spectrum of PGWs through the correlation:
\begin{align}
    \Omega_{\rm GW}(k)h^2 = 6.8\times 10^{-7} \mathcal{P}_h(k).
\end{align}

\subsection{Starobinsky's linear potential}

In this instance, the inflationary energy scale remains elusive, given that the curvaton field plays a role in the generation of primordial scalar perturbations. Consequently, $V_0$ emerges as a somewhat unconstrained parameter. To effectively explain the PTA signal through PGWs, we select the following parameters: $V_0=10^{-14}$, $A_+/V_0=1$, $A_-/A_+=0.1$, $\phi_0=6$, and $\alpha=27$. Additionally, we designate the field value of $\phi$ at the moment when the CMB scale $k_{\rm CMB}=0.05\rm{Mpc}^{-1}$ exits the horizon to be $11.32$. Moreover, we define the \textit{e}-folding number from the time when $\phi=11.32$ to the end of inflation to be $60$.

\begin{figure}[htbp]
\centering
\includegraphics[width=0.9\columnwidth ]{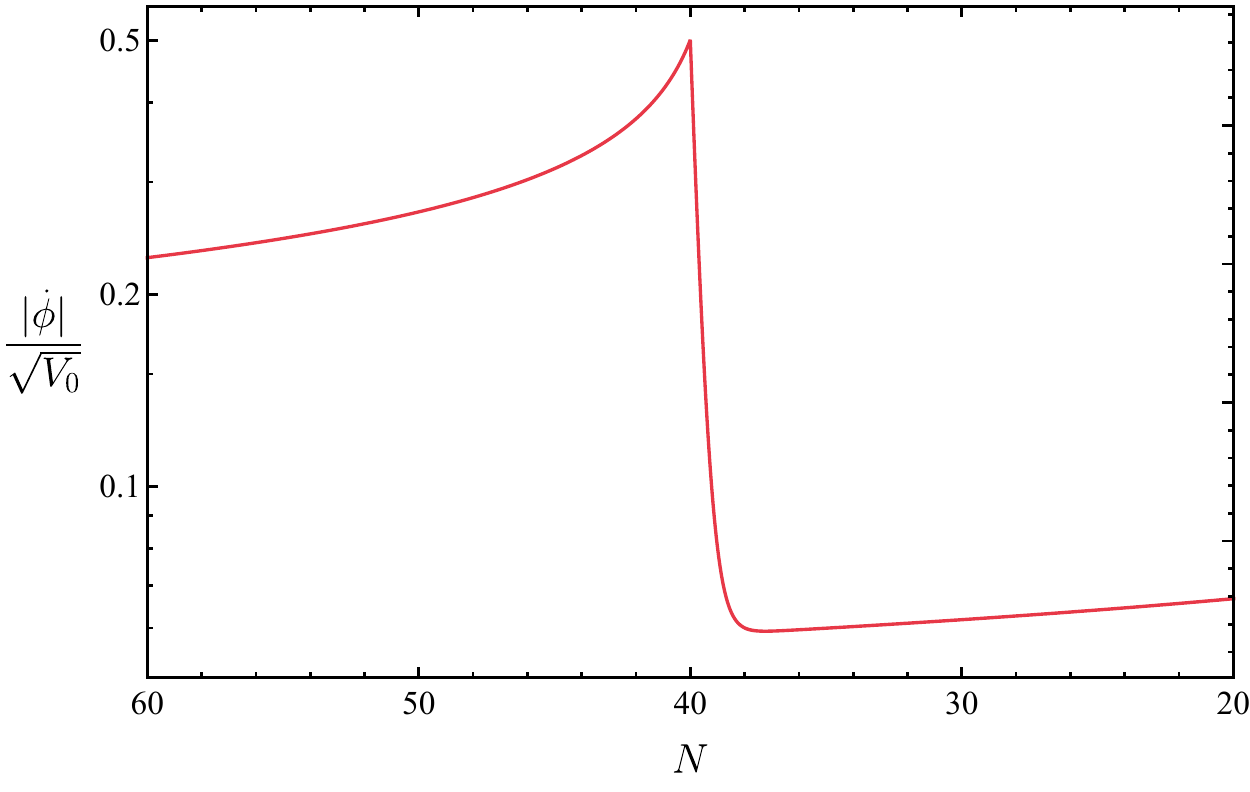}
\caption{The evolution of $|\dot\phi|/\sqrt{V_0}$ as a function of the \textit{e}-folding number $N \equiv \ln(a_{\rm e}/a)$, with $a_{\rm e}$ being the scale factor at the end of inflation, in the first scenario.}
\label{fig:phidot-N-sce1}  
\end{figure}

In Fig.~\ref{fig:phidot-N-sce1}, we depict the trajectory of the $\phi$ field velocity in relation to the \textit{e}-folding number $N$. At the outset, the inflaton $\phi$ slowly rolls down along the potential's segment where $\phi>\phi_0$, with its velocity seeing a steady uptick, culminating around $N = 40$. Subsequently, the $\phi$ field crosses the point at $\phi=\phi_0$, transitioning onto the flatter segment of the potential where $\phi<\phi_0$. In the process, the inflaton velocity declines sharply, settling into a new slow-roll regime characterized by the smaller velocity. Consequently, the trajectory of the inflaton velocity prominently features a sharp peak, manifesting itself around $N=40$.

Taking a closer look at Eq.~\eqref{eq:EoM_h_k}, it is evident that when $\dot\phi < 0$ and $\alpha>0$, the right-handed polarization state for tensor perturbations undergoes tachyonic instability. This instability translates to an exponential growth of the modes with right-handed polarization. However, this is not a uniform growth for all modes. Modes that exit the horizon at varying times will witness different levels of this instability, precisely because the strength of the tachyonic instability experienced by the modes is directly proportional to $|\alpha\dot\phi|$ evolving over time. Therefore, those modes that exit the horizon around the moment when $|\alpha\dot\phi|$ is at its largest will encounter the most pronounced amplification.

\begin{figure}[htbp]
\centering
\includegraphics[width=0.9\columnwidth ]{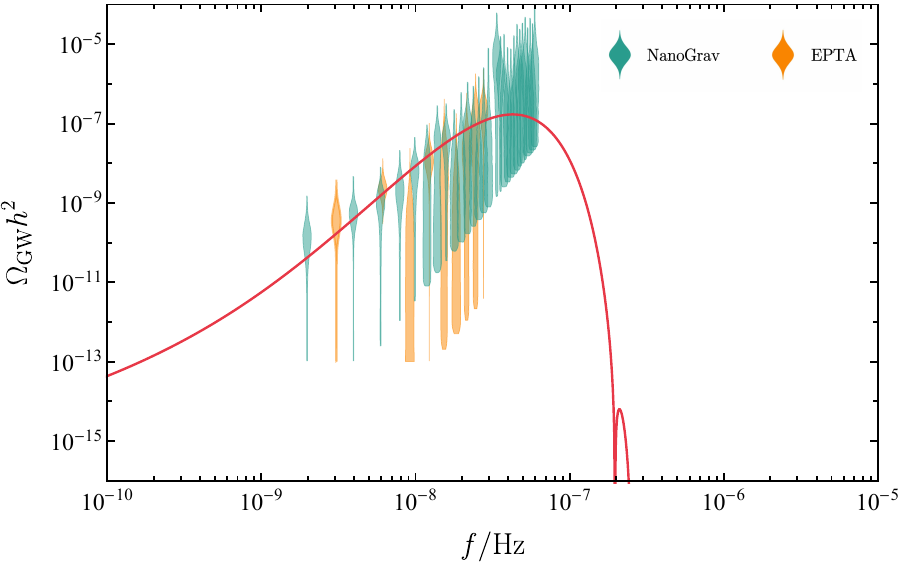}
\caption{The current energy spectrum of PGWs predicted by the NYmTG model within the first inflationary scenario. The violin diagrams depict the free spectrum posteriors in the analyses from the NANOGrav 15-yr data set \cite{NANOGrav:2023hvm} and EPTA DR2 \cite{EPTA:2023xxk}.}
\label{fig:OmegaGW-sce1}  
\end{figure}

In light of the inflationary dynamics delineated in Fig. \ref{fig:phidot-N-sce1}, the energy spectrum of the resulting PGWs exhibits a pinnacle, involving the contribution of only right-handed polarization state, in the vicinity of $\mathcal{O}(10^{-8})$Hz, as discerned from Fig. \ref{fig:OmegaGW-sce1}. From this illustration, it becomes lucidly clear that the forecasted PGW signal coincides with the observational results from NANOGrav and EPTA.

\begin{figure}[htbp]
\centering
\includegraphics[width=0.9\columnwidth ]{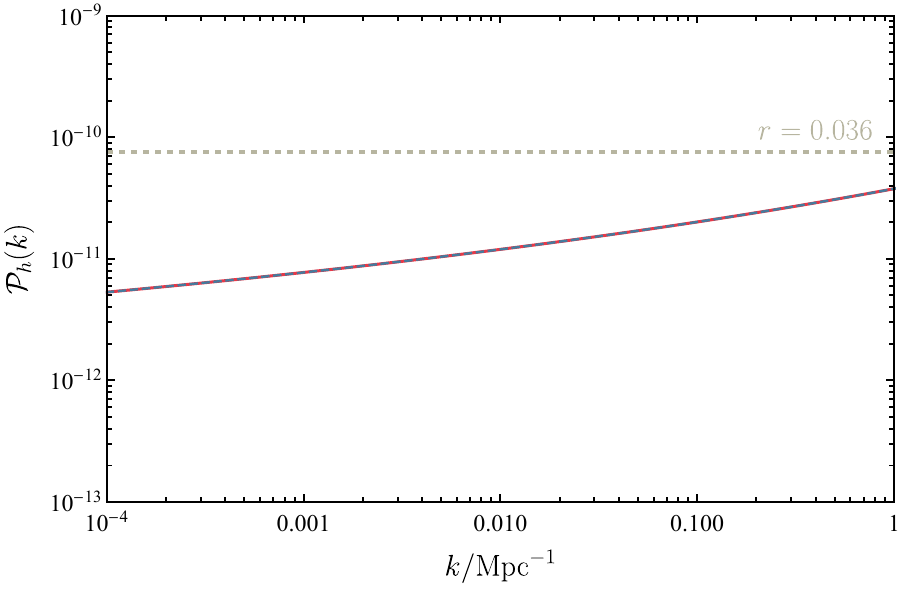}
\caption{The large-scale power spectrum of the tensor perturbations predicted by the NYmTG model within the first inflationary scenario. The solid line denotes the total power spectrum $\mathcal{P}_h(k)$, while the dashed line represents the power spectrum for the right-handed polarization. The scale-invariant tensor spectrum with $r=0.036$ (gray dashed line) is shown as a reference.}
\label{fig:spec-sce1} 
\end{figure}

\begin{figure}[htbp]
\centering
\includegraphics[width=0.9\columnwidth ]{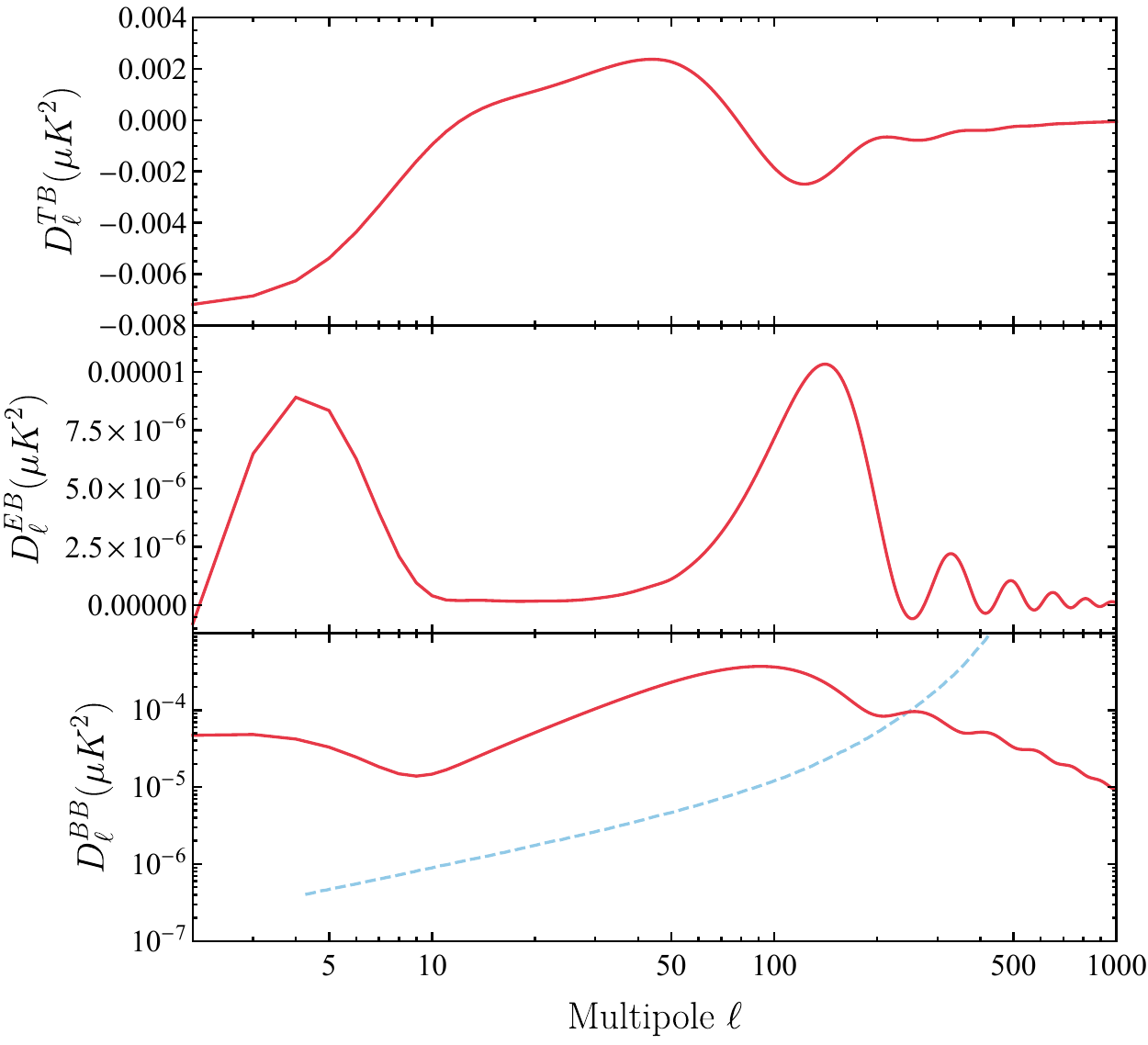}
\caption{The resulting CMB angular power spectra, $D^{XY}_\ell = \ell (\ell +1 )C^{XY}_\ell / (2\pi)$ with $XY=\{BB,EB,TB\}$, expected due to the chiral tensor spectrum depicted in Fig. \ref{fig:spec-sce1}. The dashed line in the lower panel represents the expected sensitivity curve of the LiteBIRD mission \cite{Ishino:2016izb}.}
\label{fig:spec-ang-sce1} 
\end{figure}

Intriguingly, even when considering large scales, the tensor modes with right-handed polarization exhibit distinct variations, setting them apart from their left-handed counterpart. As illustrated in Fig. \ref{fig:spec-sce1}, the large-scale tensor perturbations are characterized by a pronounced blue spectrum, within the current upper limit of $r<0.036$. More importantly, the CMB-scale PGW signal, predominantly governed by the right-handed polarization state, is nearly exclusively right-handed in its polarization. Such chiral PGWs not only yield the BB spectrum of CMB but also result in nozero TB and EB spectra. In Fig. \ref{fig:spec-ang-sce1}, we present the corresponding BB, TB, and EB spectra, derived by using the publicly available Boltzmann code CLASS~\cite{Lesgourgues:2011re}. Notably, the BB spectrum appears detectable by forthcoming CMB experiments, such as the LiteBIRD satellite \cite{Ishino:2016izb}. A detection of the TB and EB correlations would provide a strong evidence for our scenario.

\subsection{Axion potential}

In this setting, the action $S_{\rm other}$ delineates a canonical and minimally coupled inflaton, denoted by $\varphi$. 
For convenience, we adopt the Starobinsky potential, $V(\varphi) = V_0 \left[ 1 - \exp\left(- \sqrt{2/3}~\varphi\right) \right]^2$ with $V_0=9.75\times 10^{-11}$, serving as an emblematic exemplar to illustrate the results within this scenario. We opt for an initial field value of the inflaton, $\varphi=5.42$, corresponding the time when the scale $k_{\rm CMB}$ exits the horizon and ensuring that the inflation endures for a span of $60$ \textit{e}-folds. The parameters inherent to the potential, as expressed in Eq. \eqref{axion_pot}, are selected as follows: $m = \sqrt{0.16V_0}$, $\sigma = 0.0002$, and $\beta \equiv \Lambda^4/(m^2\sigma^2)=0.93$. Furthermore, we take $\alpha=1.33\times10^{5}$, and the initial field value of the $\phi$ field, denoted by $\phi_{\rm ini}$, is chosen to be $\phi_{\rm ini} = \phi_0 = 6.612 \times 10^{-4}$.

\begin{figure}[htbp]
\centering
\includegraphics[width=0.9\columnwidth ]{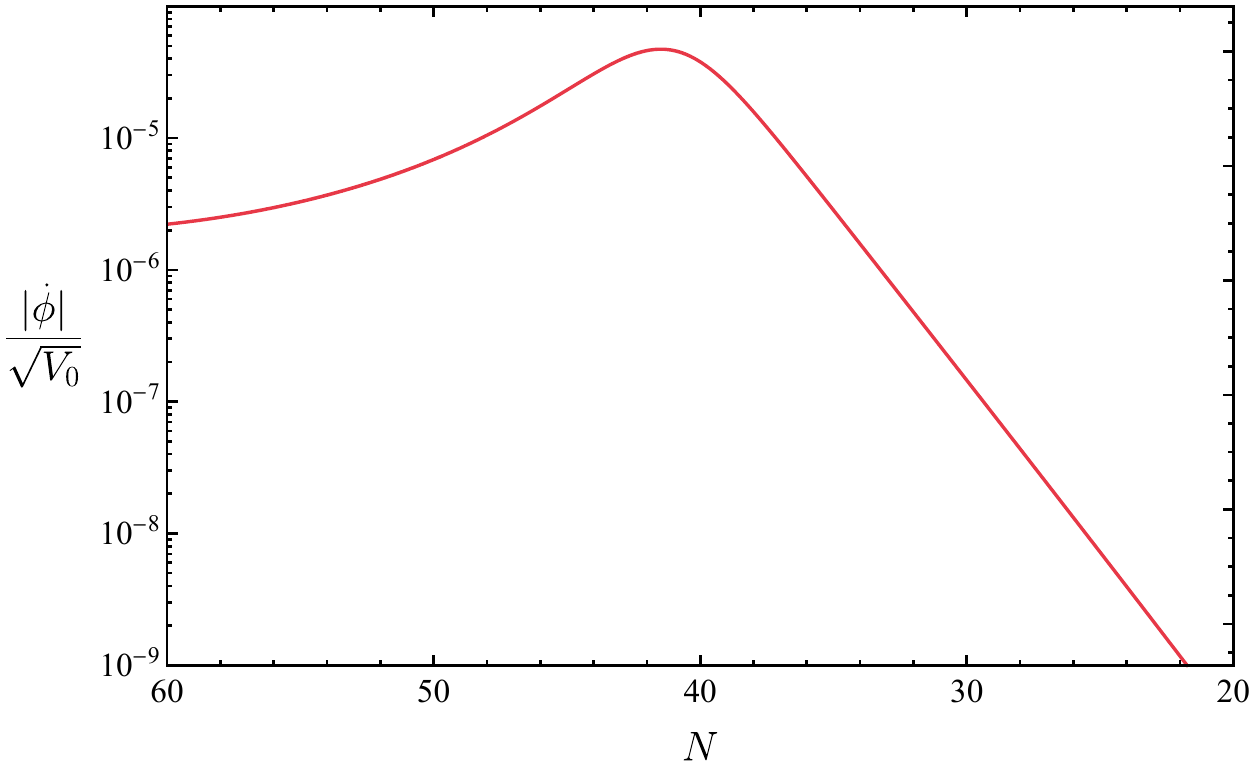}
\caption{The evolution of $|\dot\phi|/\sqrt{V_0}$ as a function of the \textit{e}-folding number $N$ in the second scenario.}
\label{fig:phidot-N-sce2}  
\end{figure}

\begin{figure}[htbp]
\centering
\includegraphics[width=0.9\columnwidth ]{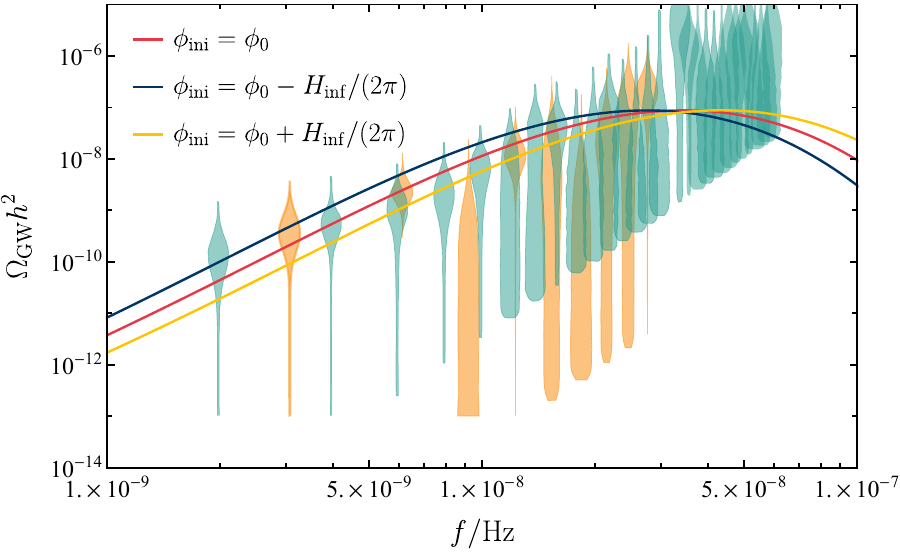}
\caption{The current energy spectrum of PGWs predicted by the NYmTG model within the second inflationary scenario. We also show the predictions under selecting different initial field values, $\phi_{\rm ini} = \phi_0 \pm H_{\rm inf}/(2\pi)$, with $H_{\rm inf}$ denoting the Hubble parameter during inflation.}
\label{fig:OmegaGW-sce2}  
\end{figure}

Analogous to the outcomes in the previous scenario, Fig. \ref{fig:phidot-N-sce2} unmistakably reveals a pronounced peak in the profile of $|\dot\phi|$ approximately at $N=40$, attributed to the swift descent of the $\phi$ field over the cliff-like region of its potential during this brief period. Consequently, the forecasted chiral PGW signal emerges as a plausible explanation of the PTA results, as seen from Fig. \ref{fig:OmegaGW-sce2}. 

The characteristic of this scenario is generating large anisotropies in the SGWB, allowing the PTA experiments to detect the GW polarization. Because of the poor angular resolution of the GW observers, we only focus on the large-scale perturbations in GW energy density which correspond to the low-$\ell$ region of the anisotropies.
In this case, quantum fluctuations of $\phi$ in the first several $e$-foldings of inflation result in large-scale inhomogeneous distribution of $\phi$. Since $\phi$ is frozen on the potential at the beginning, $\left|\frac{\dot{\phi}}{H_{\mathrm{inf}}\phi_{\mathrm{ini}}}\right|\ll 1$~($H_{\rm inf}$ denotes the Hubble parameter during inflation), superhorizon perturbations of $\phi$ remain almost constant until $\phi$ rolls down the potential. Also, superhorizon perturbations of $\phi$ can be treated as ``initial value" of the dynamics of $\phi$. As a consequence, in each large-scale region, the equivalent initial value of $\phi$ is different from each other by a quantity of the order of $H_{\rm inf}/(2\pi)$.  
Given the diminutive initial field value of $\phi$, even small fluctuations in $\phi$ can result in non-negligible difference of the time when the $\phi$ field traverses the cliff-like region in its potential. Since the peak value of amplified GWs is sensitive to this particular time, the averaged GW energy spectrum in each large-scale region differs with a shift in frequency.

\begin{figure}[htbp]
\centering
\includegraphics[width=0.9\columnwidth ]{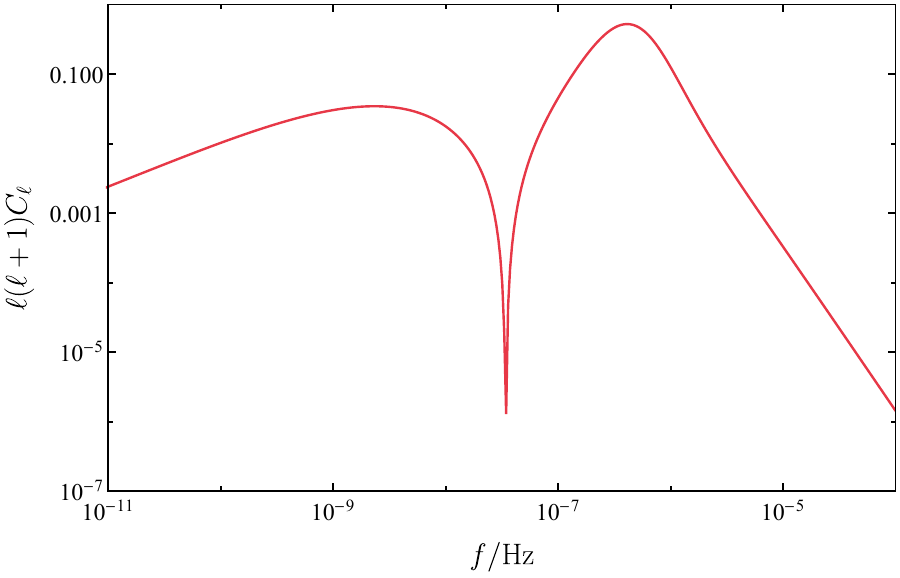}
\caption{\label{fig:spec-ang-f}  The frequency-dependent angular power spectrum predicted by the axion potential.}
\end{figure}

In Fig.~\ref{fig:OmegaGW-sce2}, we present the resulting GW energy spectra with different initial field values to demonstrate the mechanism of generating anisotropies, where the blue and yellow lines show $\Omega_{\mathrm{GW}}$ with $\phi_{\rm ini} = \phi_0 \pm H_{\rm inf}/(2\pi)$ as comparisons to the averaged case with $\phi_{\rm ini}=\phi_0$. Fig.~\ref{fig:OmegaGW-sce2} implies that the initial value from large-scale fluctuations of $\phi$ result in a frequency shift of $\Omega_{\mathrm{GW}}$. As a consequence, the value of $\Omega_{\mathrm{GW}}$ averaged in each large-scale region changes with $\phi_{\mathrm{ini}}$, i.e., anisotropies of the SGWB. Note that the anisotropies is also a frequency-dependent variable. We use the angular power spectrum to quantize the anisotropies. Similar to the Sachs-Wolfe plateau of CMB temperature
perturbations for small multiple $\ell$~\cite{Liddle:2000cg}, we obtain
\begin{equation}\label{eq:SWP}
\ell(\ell+1)C_{\ell}(f)=\frac{\pi}{2}\langle\delta\Omega_{\mathrm{GW}}^{2}(f,\mathbf{x})\rangle\,.
\end{equation}
Here, $\delta\Omega_{\mathrm{GW}}(f,\mathbf{x})\equiv (\Omega_{\mathrm{GW}}(f,\mathbf{x})-\overline{\Omega_{\mathrm{GW}}}(f))/\overline{\Omega_{\mathrm{GW}}}(f)$ denotes the spatial inhomogeneity of the GW energy spectrum, $\overline{\Omega_{\mathrm{GW}}}(f)$ is the averaged GW energy spectrum over the whole space, and the angle bracket denotes the power spectrum. Note that Eq.~\eqref{eq:SWP} is valid for any small $\ell$. As we mentioned before, the angular power spectrum depends on the frequency $f$, which is shown in Fig.~\ref{fig:spec-ang-f}. The frequency-dependent angular power spectrum of order $\mathcal{O}(0.1)$ is a distinct feature to be verified by future PTA observations. Moreover, the polarization of the SGWB can be measured by the PTA experiments thanks to the large anisotropies.

\section{Conclusion and discussion}
In this work, we examine the potential interpretation of the recently detected signal by PTA observations as the PGWs stemming from inflation within a parity-violating gravity model known as the NYmTG model. This model violates the parity symmetry in gravity by introducing the scalar field coupled Nieh-Yan term into teleparallel equivalent of general relativity, equivalently making a parity-violating extension to GR. Within the framework of the NYmTG model, we evaluate two distinct inflationary scenarios: in one, the coupled scalar field serves as an inflaton with Starobinsky's linear potential, while in the other, it acts as a spectator field governed by an axion potential. Both scenarios lead to a significant amplification of right-handed polarized tensor perturbations at certain scales, predicting a chiral PGW signal that coincides with the results from the PTA observations.
Furthermore, our results reveal that the predicted PGWs manifest distinct characteristics: for the former scenario, strong circular polarization is evident at large scales, resulting in the nonzero TB and EB spectra; for the latter scenario, significant anisotropies emerge in PTA frequency band, rendering the polarization detectable through PTA experiments. With further experimental improvements, these features would become observable, offering compelling evidence for our scenarios.
Such observations promise to shed light on both the inflationary dynamics and the underlying gravity theory.

\begin{acknowledgments}
We thank Zu-Cheng Chen and Lang Liu for fruitful discussions.
This work is supported by the National Key Research and Development Program of China Grant No. 2020YFC2201502, the National Natural Science Foundation of China Grants No. 12305057, No. 12105060, No. 12147103, No. 12235019, No. 12075297 and No. 12147103, in part by the Science Research Grants from the China Manned Space Project with NO. CMS-CSST-2021-B01,	in part by the Fundamental Research Funds for the Central Universities.
XYY is supported in part by the KIAS Individual Grant QP090701.

\end{acknowledgments}

\bibliography{references}  

\end{document}